# Nonlinear Damping of the LC Circuit using Anti-parallel Diodes

Edward H. Hellen[a)] and Matthew J. Lanctot[b)]

Department of Physics and Astronomy, University of North Carolina at Greensboro,

Greensboro, NC 27402

**Abstract**

We investigate a simple variation of the series RLC circuit in which anti-parallel diodes replace the resistor. This results in a damped harmonic oscillator with a nonlinear damping term that is maximal at zero current and decreases with an inverse current relation for currents far from zero. A set of nonlinear differential equations for the oscillator circuit is derived and integrated numerically for comparison with circuit measurements. The agreement is very good for both the transient and steady-state responses. Unlike the standard RLC circuit, the behavior of this circuit is amplitude dependent. In particular for the transient response the oscillator makes a transition from under-damped to over-damped behavior, and for the driven oscillator the resonance response becomes sharper and stronger as drive source amplitude increases. The equipment is inexpensive and common to upper level physics labs.

*Keywords*: nonlinear oscillator; LC circuit; nonlinear damping; diode

## I. Introduction

The series RLC circuit is a standard example of a damped harmonic oscillator-one of the most important dynamical systems. In this paper we investigate a simple variation of the RLC circuit in which the resistor is replaced by two anti-parallel diodes. This oscillator circuit is fundamental in the sense that it is constructed from a small number of the most basic passive electrical components: inductor, capacitor, and diodes. The result is that oscillator characteristics that have no amplitude dependence for the standard RLC circuit have strong amplitude dependence for the oscillator presented here. Transient response, voltage gain, resonant frequency, and sharpness of resonance exhibit threshold behavior due to the nonlinear current-voltage characteristics of the diodes. We show that measured behavior agrees remarkably well with numerical prediction using standard circuit component models.

The use of anti-parallel diodes creates a current-dependent damping term. Section II shows that this oscillator is described by the homogeneous nonlinear differential equation:

$$\frac{d^2 x}{d\tau^2} + (b_0 + b_1(x))\frac{dx}{d\tau} + x = 0 \qquad (1)$$

where

$$b_1(x) = \begin{cases} \dfrac{b_{max}}{|x|} & |x| \gg 1 \\ b_{max} & |x| \to 0 \end{cases}. \qquad (2)$$

Here $x$ is the dimensionless current in the circuit, $\tau$ is dimensionless time, $b_0$ is the familiar constant damping term and $b_1(x)$ is the current-dependent damping term. Thus damping is largest at the equilibrium position $x = 0$, and decreases inversely and symmetrically about it. The nonlinearity of Eq. (1) is due solely to $b_1(x)$. Our interest is in the case when damping is dominated by $b_1(x)$. This term is shown to be responsible for the amplitude dependent behavior of the oscillator.

Equation (1) is nonlinear, but when expressed as a system of first order equations is easily solved numerically using standard methods such as an adaptive Runge-Kutta. We compare these numerical predictions with experimental measurements for both the transient response and the steady-state response to a sinusoidal source.

The circuit investigated here looks similar to the extensively studied chaotic oscillator constructed from a sinusoidal signal generator driving the series combination of a resistor, inductor, and diode.[1-5] In those investigations interesting dynamics including chaos can occur when the reverse-recovery time of the diode is comparable to the circuit's natural oscillation period. For the circuit in this paper the reverse-recovery time is assumed to be zero since it is much shorter than the oscillation period. In addition we include a capacitor so that the diode's capacitance is negligible. It is the voltage dependence of the diode's dynamic resistance that causes interesting behavior.

This oscillator circuit has analogy with the "eddy-current damped" pendulum, often used as a demonstration of Faraday's and Lenz's Laws. In this demonstration a metal disk pendulum passes through the poles of a magnet placed at the bottom of the swing. Thus the damping occurs near the equilibrium position $x = 0$ when the speed $dx/dt$ is highest. For the oscillator presented here this corresponds to maximum damping at zero current and maximum voltage across the capacitor and inductor. However the analogy does not extend to the mathematical form of the position dependence of the damping.

**II. Circuit analysis**

The nonlinearly damped LC circuit using anti-parallel diodes is shown in Figure 1. Summing voltages around the circuit and then taking the time derivative gives the equation

$$L\frac{d^2I}{dt^2} + R\frac{dI}{dt} + \frac{dV_d}{dt} + \frac{1}{C}I = \frac{dV_s}{dt} \qquad (3)$$

where $V_d$ is the voltage across the anti-parallel diodes. The resistor $R$ accounts for the intrinsic resistance of the inductor in addition to any explicit resistor. For the transient response the right hand side is zero. [In actual measurements a square wave source $V_s$ makes its transition at time zero from $V_0$ to the square wave's other value $V_f$. This provides the initial conditions of zero current and $V_0$ for the capacitor voltage, and thus $dI/dt = (V_f - V_0)/L$.] For the steady state response $V_s$ is a sinusoidal source whose amplitude and frequency are varied.

The voltage $V_d$ across the anti-parallel diodes is a function of current, so its time derivative is written as

$$\frac{dV_d}{dt} = \frac{dV_d}{dI}\frac{dI}{dt} = R_d(I)\frac{dI}{dt} \qquad (4)$$

where $R_d(I) = dV_d/dI$ is the dynamic resistance of the anti-parallel diodes. This is derived as follows.

The standard current-voltage relation for a diode is

$$I(V_d) = I_0 \left[ \exp\left(\frac{eV_d}{mkT}\right) - 1 \right] \tag{5}$$

where $e$ is the elementary charge, $k$ is Boltzmann's constant, $T$ is the absolute temperature, and $V_d$ is the voltage across the diode.[6] At room temperature $kT/e = V_{th} \approx 25$ mV. Quantity $I_0$ is the reverse saturation current and $m$ is a correction factor. Typical values for a silicon diode are $m = 2$ and a few nanoamperes for $I_0$. Applying Eq. (5) to anti-parallel diodes as in Fig. 1 results in the current-voltage relation

$$I(V_d) = 2I_0 \sinh\left(\frac{V_d}{mV_{th}}\right). \tag{6}$$

We define dimensionless quantities for voltage and current:

$$v = \frac{V_d}{mV_{th}} \qquad i = \frac{I}{2I_0}. \tag{7}$$

The current-voltage relation and the dynamic resistance for the anti-parallel diodes are then

$$i(v) = \sinh(v) \quad v(i) = \sinh^{-1}(i) \tag{8}$$

$$R_d(i) = \frac{dV_d}{dI} = \frac{mV_{th}}{2I_0}\frac{dv}{di} = \frac{mV_{th}}{2I_0 \cosh(v(i))} = \frac{R_{max}}{\cosh(\sinh^{-1}(i))}. \tag{9}$$

Putting in values for $m$, $V_{th}$, and $I_0 = 3$ nA gives $R_{max} = mV_{th}/(2I_0) = 8.3\,M\Omega$. Figure 2 shows the dynamic resistance, Eq. (9), as a function of dimensionless current. The behavior of the dynamic resistance in Eq. (9) for large and small currents is

$$R_d(i) = \begin{cases} \dfrac{R_{max}}{|i|} & |i| \gg 1 \\ R_{max} & i \to 0 \end{cases}. \quad (10)$$

The inverse current relationship for the dynamic resistance is generally known for a conducting diode ($|i| \gg 1$), usually stated as $25/I$ $\Omega$ (where $I$ is in milliamperes and $m = 1$ for an ideal diode).[7] The inverse current relationship is shown as dashed lines in Fig. 2.

We define dimensionless time $\tau = t/\sqrt{LC}$ and using Eq. (4) rewrite Eq. (3) for the transient response in terms of dimensionless quantities:

$$\frac{d^2 i}{d\tau^2} + \left( \sqrt{\frac{C}{L}} R + \sqrt{\frac{C}{L}} R_d(i) \right) \frac{di}{d\tau} + i = 0 . \quad (11)$$

Thus we have shown that the oscillator in Fig. 1 is described by Eqs. (1) and (2) where

$$b_0 = R\sqrt{\frac{C}{L}} \quad b_1(x) = \sqrt{\frac{C}{L}} R_d(i) \quad b_{max} = R_{max}\sqrt{\frac{C}{L}}. \quad (12)$$

### III. Procedure

#### A. Numerical Computation

For the numerical solution of the equations we use a system of first order nonlinear differential equations for the current $I$ and the voltage across the capacitor $V_C$. Application of Kirchhoff's voltage law to Fig. 1 gives;

$$\frac{dI}{dt} = \frac{-RI}{L} - \frac{mV_{th}}{L} \sinh^{-1}\left(\frac{I}{2I_0}\right) - \frac{V_C}{L} + \frac{V_S(t)}{L} \quad (13a)$$

$$\frac{dV_C}{dt} = \frac{I}{C}. \quad (13b)$$

This system of equations is computed numerically using an adaptive Runge-Kutta method in Matlab.

For circuit measurements of the transient response we use a square wave with frequency set low enough (~ 1 Hz) so that each transition of the square wave gives a well-defined initial voltage $V_0$ and final voltage $V_f$ for the capacitor voltage. Thus for the numerical solution of the transient response $V_S(t)$ is $V_f$ and the initial conditions are zero current and $V_0$ for the capacitor voltage.

For the steady state response $V_S(t)$ is a sinusoidal function. Solutions are computed for different amplitudes and frequencies of source $V_S(t)$. The amplitude of the calculated capacitor voltage is recorded after the transient response has died off so that the solution has clearly achieved steady-state. The voltage gain $A$ of the circuit is the ratio of the capacitor voltage amplitude to the source amplitude. The results of the numerical computations are predictions for how the gain, the sharpness of the resonance, and the resonant frequency depend on the source amplitude.

**B. Circuit Construction and Data Collection**

For the oscillator circuit we used a 2.5 mH inductor with an intrinsic resistance of 9 Ω, a 0.010 μF capacitance, and two 1N4148 silicon diodes. The resulting values for the dimensionless damping terms in Eq. (12) are $b_0 = 0.018$ and $b_{max} = 16600$. The diodes are characterized by $m = 2$ and $I_0 = 3 \times 10^{-9}$ amp. These values were obtained by measuring the diode's current-voltage characteristics and fitting to Eq. (5). They agree well with previously measured values.[8] This diode model was used for numerical predictions of all the experiments. However it is known that Eq. (5) deviates from the measured I-V curve when used for both small and large currents.[6] Therefore, for experiments involving large currents an additional numerical prediction was made that included a second set of parameters, $m = 2.76$ and $I_0 = 1.7 \times 10^{-7}$ amps, for currents greater than 5 ma.

The inductance and capacitance give a natural oscillation frequency of 31.8 kHz and corresponding period of about 30 µs. This is much larger than the 1N4148's reverse-recovery time of less than 4 ns (from datasheet). Thus the reverse-recovery time is assumed to be zero.

The capacitance was comprised of 5 polypropylene nominal 0.002 µf capacitors in parallel. As discussed later, this was done in order to minimize the effect of the capacitor's equivalent series resistance (ESR). Thus the total resistance in the circuit is attributed to the inductor's intrinsic resistance of 9 Ω.

The transient response is measured by using a source $V_s$ with a low frequency (~1 Hz) square wave, thus providing transitions between the two values of the square wave. The quality of the square wave is important. The transition should be complete in a time much shorter than the natural oscillation period of about 30 µs. A square wave from a typical signal generator will not be able to drive the circuit due to the relatively high output impedance of the signal generator. Therefore we run the signal generator to the simple drive circuit shown in Fig. 3. The result is a square wave with transitions completed within about 1 μs. Generic op amps and small signal transistors are sufficient. If desired, a common slightly more complicated driver circuit that uses two conducting diodes to separate the transistor bases may be used to eliminate cross-over glitches.

For the steady-state measurements the signal generator produced a sinusoidal voltage with the desired amplitude and frequency. For investigations of how the maximum gain $A_{max}$ depends on source amplitude, the frequency was adjusted to maximize the capacitor voltage for each source amplitude. This procedure also provided data for how the resonant frequency depends on source amplitude. For investigations of the sharpness of the resonance response, the capacitor voltage was measured as a function of frequency at two source amplitudes.

Data were collected on a Tektronix TDS 1002 digital oscilloscope. A standard 10 MΩ oscilloscope probe attached to the capacitor will affect the circuit's behavior. This is because 10 MΩ is comparable to the dynamic resistance in Fig. 2 for small currents. Therefore the capacitor voltage was buffered by a unity gain FET input op amp (LF411 for example) before measurement by the oscilloscope.

### C. Equivalent Series Resistance (ESR)

It is important to consider the capacitor's ESR. Use of a general purpose ceramic disk capacitor without including its ESR will result in poor agreement between measurement and prediction. We chose to minimize the ESR by using a parallel combination of 5 polypropylene capacitors. The ESR of this combination appeared to be no more than 1Ω.

Using a 0.01 µF ceramic disk capacitor instead of the polypropylene capacitors required adding 15 Ω to the inductor's 9 Ω in order to get good agreement between measurement and numerical prediction. Thus the ESR of the ceramic capacitor is about 15 Ω. It's predicted ESR is obtained by using the dissipation factor from the capacitor data sheet stated as less than 0.03, and the formula for dissipation factor:[9]

$$DF = R_{esr}\, \omega C. \qquad (14)$$

At the natural oscillation frequency of approximately 30 kHz, the predicted ESR for the 0.01 µf ceramic capacitor is thus 16 Ω. This compares well with our finding.

## IV. Results and Discussion

### A. Transient Response

Figure 4 shows data and the numerical prediction for the transient response of the voltage across the capacitor. Agreement is remarkably good. The square wave source made a transition from 8.2 to 0 volts at time $t = 0$ sec. Thus the initial conditions used in the numerical solution of Eq. (13) were zero current and 8.2 volts across the capacitor.

It is apparent that this oscillator makes a transition from under-damped to over-damped behavior when the amplitude (peak capacitor voltage) decreases below some threshold value. In Fig. 4 the last zero crossing occurs at 0.075 ms. For times after this the voltage reaches a maximum, then undergoes monotonic decay characteristic of an over-damped oscillator.

The transition from under-damped to over-damped behavior may be understood by considering the transient response oscillation frequency for a standard RLC circuit:

$$\omega = \sqrt{\frac{1}{LC} - \left(\frac{R}{2L}\right)^2} \quad \text{(standard RLC)}. \tag{15}$$

For a fixed value of $R$ the frequency is constant so there is no transition from under-damped to over-damped behavior for the standard RLC circuit. However for the anti-parallel diodes the resistance depends on the current as shown in Fig. 2. If $R$ represents an effective resistance during a complete oscillation, then Eq. (15) does predict dependence of $\omega$ on oscillation amplitude. As the amplitude decreases, the current spends more time near small values where the dynamic resistance of the anti-parallel diodes is large, causing the effective resistance to increase. Eventually the effective resistance is large enough so that $\omega$ becomes imaginary, corresponding to the transition to over-damped decay. The concept of an effective resistance is useful for understanding the observed behavior, however it is not intended as a calculation tool.

The square wave's value of zero volts [used for $V_s(t)$ in Eq. (13)] is not reached during the time shown in Fig. 4. This is due to the large dynamic resistance of the diodes at small currents. The RC time constant for small currents is $(8M\Omega) \times (0.01\mu f) \approx 0.1s$. Previously we showed that the voltage on a capacitor discharging through a diode has a nearly logarithmic decay.[8] As a result it takes about 0.5 s for the 0.01 µf capacitor to be essentially discharged. This necessitated the use of a low frequency (~ 1 Hz) square wave, thereby allowing the transient response to die out before the next transition of the square wave.

**B. Steady state response**

The characteristics of interest for the steady-state driven oscillator are the voltage gain $A$, the resonant frequency, and the sharpness of the resonance. Here we show how these characteristics for the anti-parallel diodes-LC oscillator depend on the amplitude of the driving source. In contrast, the standard RLC circuit is independent of source amplitude for these characteristics. As in the previous section, we will use the standard RLC results and the concept of an effective resistance to understand the amplitude dependent behavior of the anti-parallel diodes-LC circuit.

Figures 5 and 6 show results for how the maximum gain $A_{max}$ and the resonance frequency $\omega_{res}$ depend on source amplitude. In both graphs numerical predictions are shown by lines and results from circuit measurements by circles. To understand the variation of $A_{max}$ and $\omega_{res}$ with source amplitude consider the theoretical voltage gain $A$ for a standard RLC circuit:

$$A = \frac{1}{LC\sqrt{\left(\omega \frac{R}{L}\right)^2 + \left(\omega^2 - \frac{1}{LC}\right)^2}} \quad \text{(standard RLC)}. \tag{16}$$

The maximum gain occurs at the resonance frequency

$$\omega_{res} = \sqrt{\frac{1}{LC} - \frac{1}{2}\left(\frac{R}{L}\right)^2} \quad \text{(standard RLC)}. \tag{17}$$

Evaluating Eq. (16) at resonance gives the maximum gain

$$A_{max} = \frac{1}{RC\sqrt{\frac{1}{LC} - \left(\frac{R}{2L}\right)^2}} \quad \text{(standard RLC)}. \tag{18}$$

Clearly $A_{max}$ and $\omega_{res}$ do not depend on source amplitude for the standard RLC circuit. However if $R$ represents an effective resistance of the anti-parallel diodes, then Eqs. (17) and (18) do predict dependence on source amplitude. Small source amplitudes result in small currents which according to Fig. 2 have the largest resistance, whereas for large source amplitudes part of the cycle has large currents where the dynamic resistance drops inversely with current resulting in a decreased effective resistance. Thus Eq. (18) predicts the behavior seen in Fig. 5, $A_{max}$ increases with amplitude. In Figure 6 the resonant frequency increases with source amplitude, leveling off at 31.8 kHz, the frequency predicted for undamped oscillation using the values C = 0.01 µF and L = 2.5 mH. This behavior is predicted by Eq. (17). Equation (17) also predicts existence of the cut-off amplitude seen in Fig. 6 below which the effective resistance is large enough so the frequency becomes imaginary. This corresponds to the transition of the circuit to a low pass filter and therefore peak response at $\omega = 0$.

Figure 5 shows three different numerical predictions. The solid line is the prediction using the single parameter set for the diode I-V characteristics as described in Section III. The discrepancy between this prediction and the experimental result is accounted for by two effects; deviation from ideal diode behavior given by Eq. (5), and the ESR of the capacitance. The dashed line is the prediction when an additional set of diode parameters was included for high currents as described in Section III. Furthermore, inclusion of 1 Ω

14for the capacitor ESR and other unaccounted for distributed resistance brings the numerically predicted ratios (dot-dashed line) in agreement with the measured results.

Note that the transient response in Fig. 4 showed little discrepancy between numerical prediction and measurement. This is because the largest amplitude of capacitor voltage was about 8 volts, which in Fig. 5 is achieved with source amplitude of about 1.1 volts. The discrepancy is small for source amplitudes below this value.

Figure 7 shows the resonance response of the voltage gain for two different source amplitudes, 0.9 (lower curves) and 1.2 (upper curves) volts. Numerically predicted gains are solid lines and measured gains are symbols. The 1.2 volt numerical predictions used the 2-part parameterization for the diodes corresponding to the dashed line prediction in Fig. 5.

The general shape of these curves is that of Eq. (16) for different $R$. The sharpness of the response clearly depends on the source amplitude. The 1.2 volt source amplitude produces a much taller and narrower peak than the 0.9 volt source. This is due to the lower effective resistance of the diodes for the larger amplitude currents. In contrast the standard RLC circuit has the same shape resonance peak for all source amplitudes. Also the resonance curve in Fig. 7 is narrower for a given maximum gain than the corresponding curve for a standard RLC circuit. This is because as the frequency moves away from resonance, the current amplitude through the diodes drops causing the effective resistance to increase. Equation (16) shows that increasing $R$ is an additional contribution to reducing *A*. The voltage gain peak disappears when the source amplitude is reduced to about 0.8 volts.  For smaller amplitudes the graph is that of a low pass filter.

It is worth noting the counter-intuitive result that Eq. (17) is mathematically different from Eq. (15), the natural oscillation frequency for the transient response of the homogeneous equation.  Confusion between these formulas is an easy source of error.

**V. Conclusion**

We have shown that the fundamental electronic oscillator obtained by replacing the resistor in a series RLC circuit by anti-parallel diodes has interesting amplitude-dependent behavior not exhibited by the RLC circuit.  In particular the transient response displays a transition from under-damped to over-damped behavior, and for the driven oscillator there is threshold behavior in which the resonance response gets sharper and stronger as the source amplitude increases.  The differential equations derived for this nonlinearly damped harmonic oscillator are easily solved numerically.  Numerical predictions agree remarkably well with measured behavior.

**Acknowledgments**

This research was supported by an award from the Research Corporation.  M.L. was supported in part by the Undergraduate Research Assistantship Program at UNCG.  We thank David Birnbaum for valuable suggestions.

[a] Electronic mail:  ehhellen@uncg.edu

[b] Current address. Columbia University, Department of Applied Physics and Applied Mathematics, New York NY 10027.

**Figure Captions**

Fig. 1. The nonlinear oscillator obtained from anti-parallel diodes, an inductor L, and capacitor C. Resistor R is included to account for inherent resistance, mostly from the inductor. A voltage source $V_s$ is also included.

Fig. 2. Calculated dynamic resistance (solid line) for anti-parallel diodes using Eq. (9). Also shown is the well known inverse current approximation (dashed line) useful for currents far from zero. The dimensionless current $i = I/(2I_0)$ where $I$ is the current in amperes and $I_0$ is the reverse saturation current of the diode.

Fig. 3. Driver circuit used between signal generator and the oscillator circuit. Uses LF412 op amp and 2N3904 and 2N3906 transistors.

Fig. 4. Transient response of the capacitor voltage for source $V_s$ making transition from 8.2 volts to 0 volts at time zero. Numerical prediction (smooth line) and measured response (noisy line) are nearly indistinguishable.

Fig. 5. Maximum voltage gain $A_{max}$ as a function of source amplitude. $A_{max}$ is the ratio of voltage amplitudes of capacitor and source at the resonant frequency. Numerical predictions are lines and results from measurements are circles.

Fig. 6. Resonant frequency as a function of source amplitude. Resonance was identified by adjusting the frequency to maximize the capacitor voltage amplitude for each source amplitude. Numerical predictions are lines and measured results are circles.

Fig. 7. Resonance response of voltage gain $A$ as function of source frequency for 2 source amplitudes: 0.9 (lower) and 1.2 (upper) amps. Numerical predictions are lines and measured results are circles.

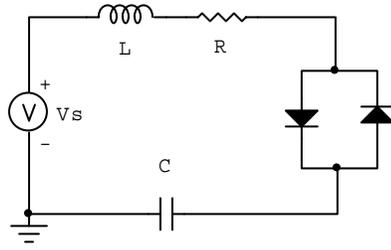

Figure 1

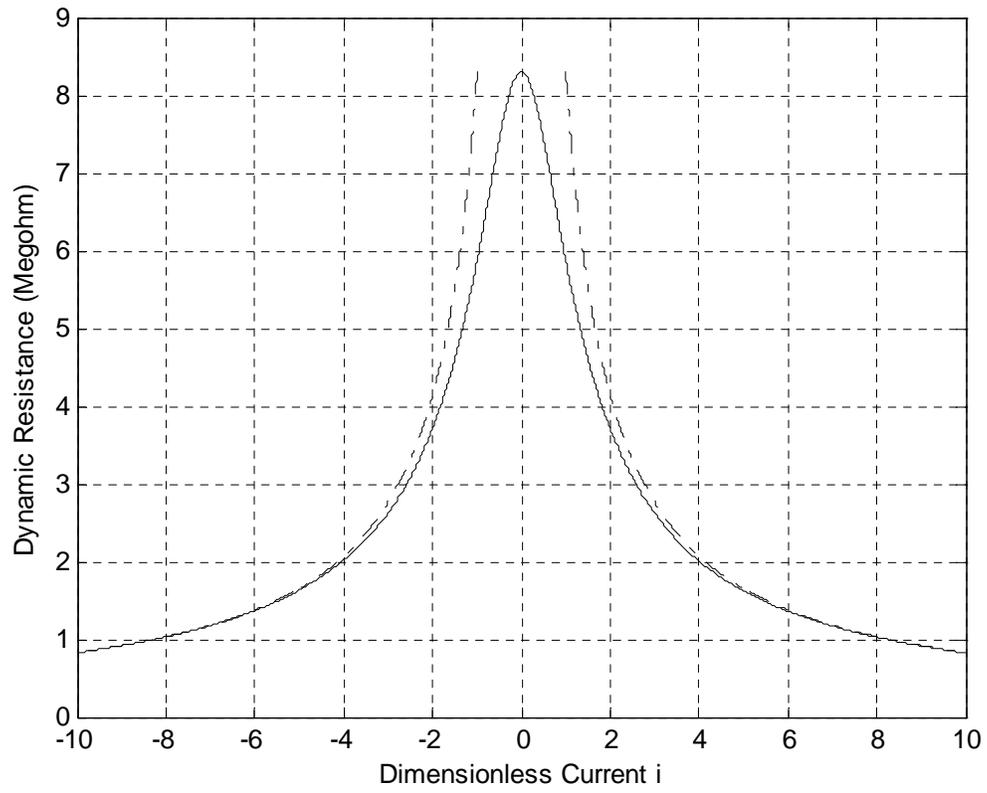

Figure 2

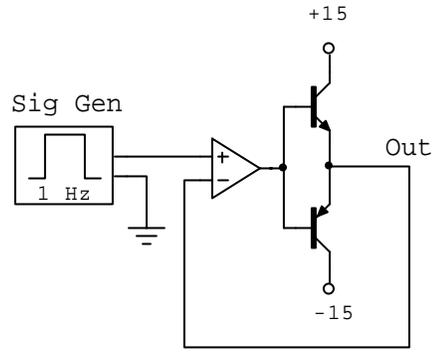

Figure 3

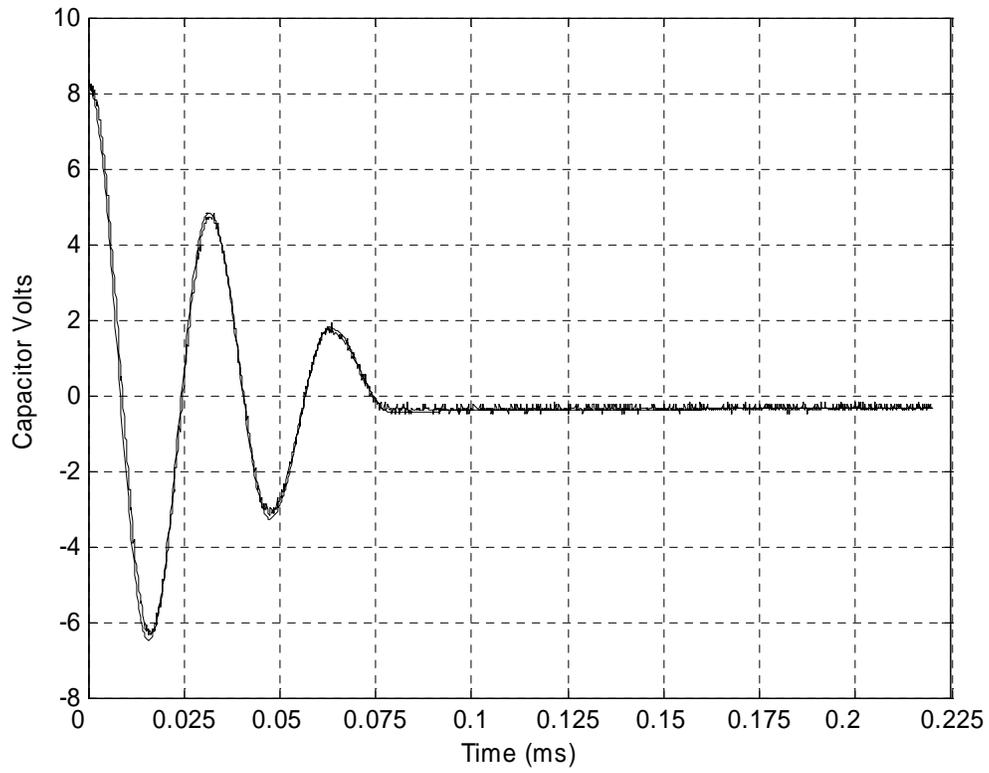

Figure 4

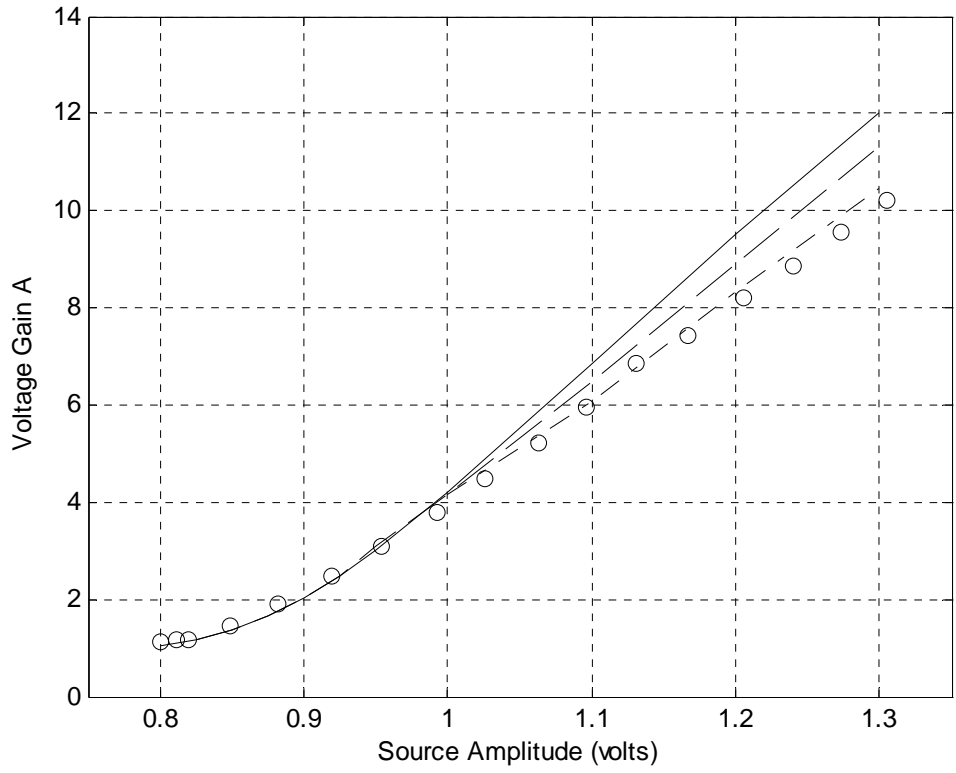

Figure 5

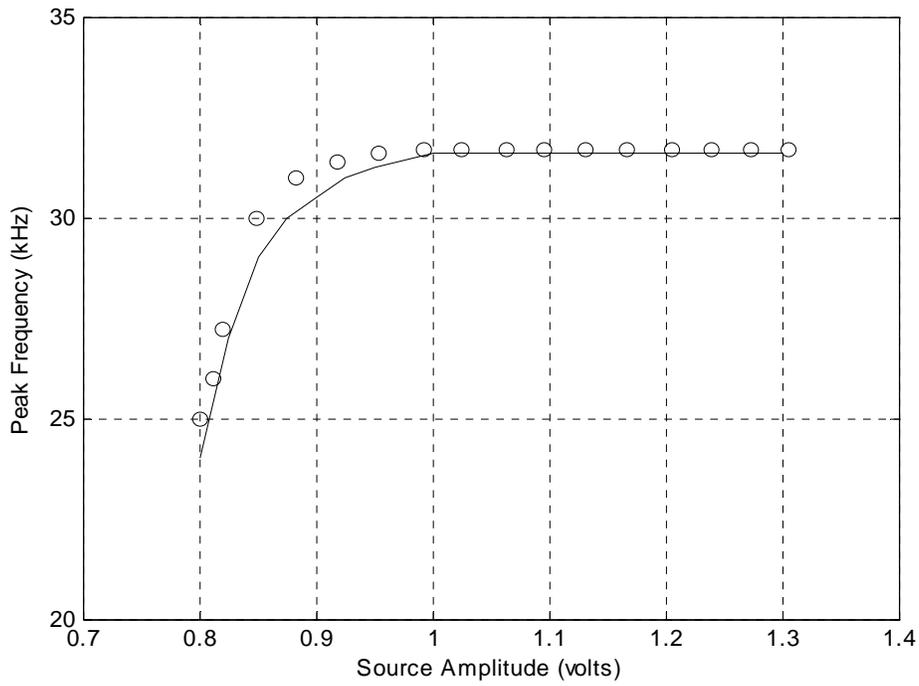

Figure 6

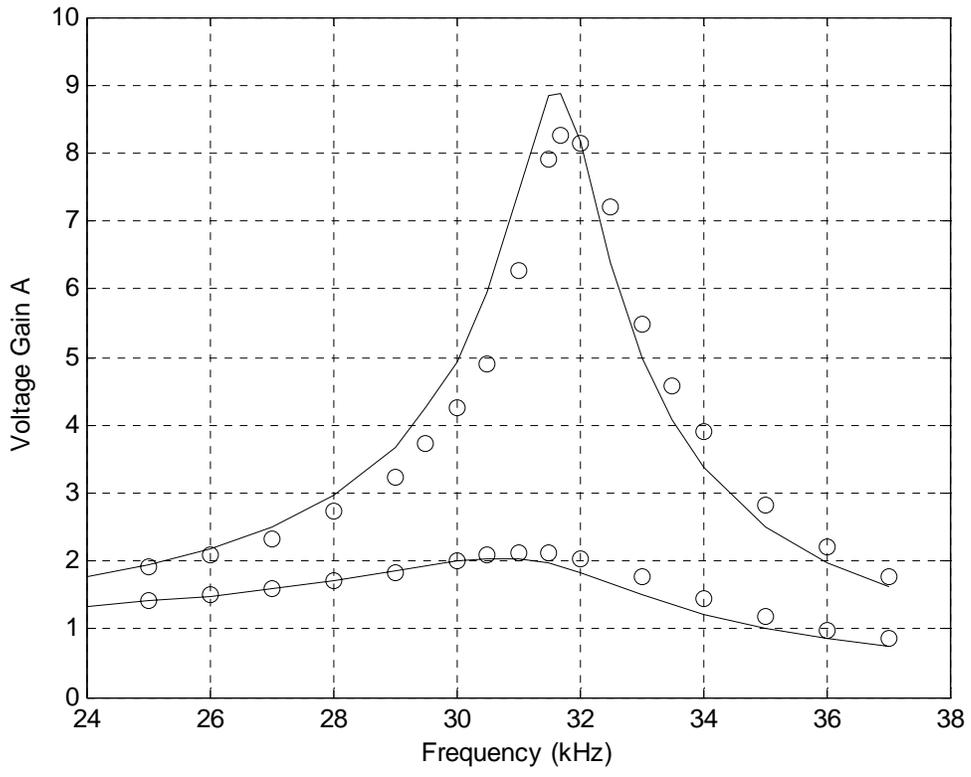

Figure 7